\documentclass[prd,a4paper,showpacs]{revtex4}

\usepackage{epsfig,epsf,graphics,psfrag}
\usepackage{times}
\usepackage{float}
\usepackage{color}
\usepackage{amsfonts,amssymb,stmaryrd,latexsym,amsmath}
\usepackage{multirow}
\usepackage{array} 
\usepackage{enumerate}

\begin{document}

\bibliographystyle{unsrt}

\title{Resolving the puzzling decay patterns of charged $Z_c$ and $Z_b$ states}


\author{Xiao-Hai Liu$^{1}$}\email{liuxiaohai@pku.edu.cn}
\author{Li Ma$^{1}$}\email{lima@pku.edu.cn}
\author{Li-Ping Sun$^{1}$}\email{sunliping@pku.edu.cn}
\author{Xiang Liu$^{2,3}$}\email{xiangliu@lzu.edu.cn}
\author{Shi-Lin Zhu$^{1,4}$}\email{zhusl@pku.edu.cn}

\affiliation{
$^1$Department of Physics and State Key Laboratory of Nuclear Physics and Technology and Center of High Energy Physics, Peking University, Beijing 100871, China\\
$^2$Research Center for Hadron and CSR Physics, Lanzhou University and Institute of Modern Physics of CAS, Lanzhou 730000, China\\
$^3$School of Physical Science and Technology, Lanzhou University, Lanzhou 730000, China\\
$^4$Collaborative Innovation Center of Quantum Matter, Beijing
100871, China}


\date{\today}

\begin{abstract}

We investigate the ratio of the branching fractions of the molecular
candidates decaying into the ground and radially excited states
within the quark interchange model. Our numerical results suggest
that these molecular candidates are more likely to decay into the
radially excited states than into ground states. Especially, the
ratio $\Gamma[Z_c(4430)\to \pi\psi(2S)]$/$\Gamma[Z_c(4430)\to \pi
J/\psi]\sim 9.8$ is close to the experimental measurement, which
supports the interpretation of $Z_c(4430)$ as the $\bar{D}D^*(2S)$
molecular state. The ratios of the branching fractions of
$Z_b(10610)$ and $Z_b(10650)$ to $\pi\Upsilon(2S, 3S)$ and
$\pi\Upsilon(1S)$ agrees very well with the Belle's measurement. We also
predict the similar ratios for $Z_c(3900)$, $Z_c(4020)$,
$R_{Z_c(3900)}$$\approx$1.3 and $R_{Z_c(4020)}$$\approx$$4.7$.
Hopefully the $\pi\psi(2S)$ mode, and ratios $R_{Z_c(3900)}$ and
$R_{Z_c(4020)}$ will be measured by the BESIII and Belle
collaborations in the near future, which shall be very helpful to
understand the underlying dynamics of these exotic states.


\end{abstract}

\maketitle

\section{Introduction}
The first evidence of the existence of charged charmonium-like
states was reported by the Belle Collaboration in
2007~\cite{Choi:2007wga}. In $B\to K \pi^{\pm}\psi(2S)$ decays, a
resonance-like structure $Z_c(4430)$ in the $\psi(2S) \pi^{\pm}$
mass distribution was observed. In 2013, the BESIII Collaboration
observed $Z_c(3900)$ in the invariant mass spectrum of $J/\psi
\pi^{\pm}$ from $Y(4260)$$\to$$J/\psi\pi\pi$~\cite{Ablikim:2013mio}.
This observation was then quickly confirmed by the Belle Collaboration
\cite{Liu:2013dau}. Another charged charged charmonium-like state
$Z_c(4020)$ ($Z_c(4025)$) was observed by BESIII in the $h_c\pi^\pm$
and $(D^*\bar{D}^*)^{\pm}$ mass distributions
\cite{Ablikim:2013wzq,Ablikim:2013emm,Ablikim:2013xfr}. There are
also two charged bottomnium-like states $Z_b(10610)$ and
$Z_b(10650)$ observed by Belle in the $e^+e^-$ annihilation into the
hidden-bottom dipion channels~\cite{Belle:2011aa}. These states are
of particular interest since their minimal quark contents are either
$c\bar{c}d\bar{u}$/$c\bar{c}u\bar{d}$ or
$b\bar{b}d\bar{u}$/$b\bar{b}u\bar{d}$, which are unambiguously
beyond the conventional $q\bar{q}$ model.

Since its observation, $Z_c(4430)$ has attracted much attention both
in experimental and theoretical
research~\cite{Rosner:2007mu,Meng:2007fu,Lee:2007gs,Liu:2007bf,Ding:2008mp,Braaten:2007xw,Branz:2010sh,Liu:2008qx,Ebert:2008kb,Dubynskiy:2008mq,Wang:2014vha}.
Because the mass of $Z_c(4430)$ is very close to the threshold of
$D_1^{(\prime)}\bar{D}^*$, the interpretation of $Z_c(4430)$ as the
loosely bound S-wave molecular state composed of
$D_1^{(\prime)}\bar{D}^*$ had been quite popular among the various
theoretical descriptions. Within this scheme, its possible
spin-parity would be $J^P$$=$$0^-$, $1^-$ or $2^-$. However, after
analyzing the $B^0\to K^+ \pi^{-}\psi(2S)$ decays, the LHCb
Collaboration not only confirmed the existence of $Z_c(4430)$ but
also determined its spin-parity to be $J^P$$=$$1^+$
unambiguously~\cite{Aaij:2014jqa}. This measurement is obviously
inconsistent with the $D_1^{(\prime)}\bar{D}^*$ molecular state
theory, which leads to very puzzling new challenge in understanding
the intrinsic structure of $Z_c(4430)$.

Very recently, it was proposed in Ref.~\cite{Ma:2014zua} that
$Z_c(4430)$ is the S-wave molecule composed of $\bar{D}D^*(2600)$ or
$\bar{D}^* D(2550) $, where the dominant components of $D^*(2600)$
and $D(2550)$ are generally accepted as the radially excited states
$D^*(2S)$ and $D(2S)$ respectively
\cite{Becher:2012xr,Wang:2013tka,Segovia:2013sxa}. In other words,
$Z_c(4430)$ may be the cousin of the charged states $Z_c(3900)$ and
$Z_c(4020)$, which are speculated to be the molecular candidates
composed of the $D$ and $D^*$ mesons. Several reasons lead to the
above ansatz. Firstly, the thresholds of $\bar{D}D^*(2600)$
($\sim$4477 MeV) and $\bar{D}^* D(2550)$ ($\sim$4546 MeV),
especially the $\bar{D}D^*(2600)$ meson pair, are very close to the
mass of $Z_c(4430)$, which is
$M_{Z_c(4430)}=$4475$\pm$7$^{+15}_{-25}$ MeV as measured by LHCb.
Furthermore, the spin-parity of the S-wave $\bar{D}D^*(2600)$
combination is $J^P$$=$$1^+$, which is also consistent with the LHCb
measurement. Since the width of $Z_c(4430)$ is very large
($\Gamma_{Z_c(4430)}$$=$172$\pm$13$^{+37}_{-34}$ MeV
~\cite{Aaij:2014jqa}), which is much larger than the width of
$D^*(2600)$ ($\Gamma_{D^*(2600)}$$=$93$\pm$6$\pm$13 MeV
\cite{Beringer:1900zz}), such a molecular state assumption is still
reasonable.

Another important motivation is due to the puzzling decay pattern of
$Z_c(4430)$. It seems that $Z_c(4430)$ is more likely to decay into
$\psi(2S)\pi$ than into $J/\psi\pi$, i.e., the ratio
$Br(Z_c(4430)^+\to\psi(2S)\pi^+)$$/$$Br(Z_c(4430)^+\to J/\psi\pi^+)$
is about 10 as measured by Belle~\cite{shencp:2014}. If $Z_c(4430)$
contains $D(2S)$ or $D^*(2S)$ as its component, one would expect
that it will decay into the final state containing a radial
excitation easily, i.e., the $\psi(2S)\pi$ channel may be its
favorable decay mode.

In this paper we will investigate the difference between the
$J/\psi\pi$ and $\psi(2S)\pi$ decay channels of $Z_c(4430)$,
$Z_c(3900)$ and $Z_c(4020)$. The calculation of the pertinent
branching ratios is mainly based on a non-relativistic
quark-interchange model. The $\Upsilon(nS)\pi$ decay modes of the
charged bottomnium-like states $Z_b(10610)$ and  $Z_b(10650)$ will
also be studied.

\section{The Model}
\begin{figure}[tb]
  \centering
  \includegraphics[width=0.5\hsize]{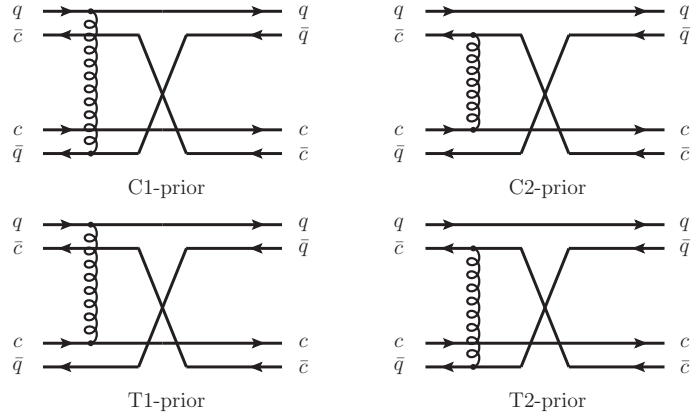}\\
  \caption{Prior quark-interchange diagrams contributing to the anticharmed meson-charmed meson scattering into pion and charmonia.}\label{priordiagram}
\end{figure}

\begin{figure}[tb]
  \centering
  \includegraphics[width=0.5\hsize]{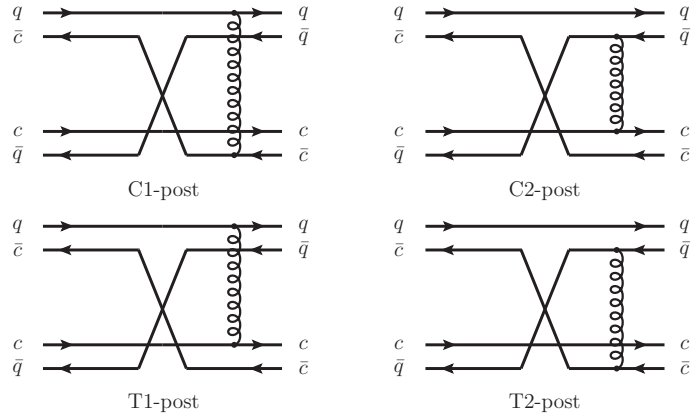}\\
  \caption{Post quark-interchange diagrams contributing to the anticharmed meson-charmed meson scattering into pion and charmonia.}\label{postdiagram}
\end{figure}

For the loosely bound S-wave molecular candidates $Z_c(4430)$,
$Z_c(3900)$ and $Z_c(4020)$ decaying into $J/\psi\pi$ and
$\psi(2S)\pi$, we can describe these decay processes as the
anticharmed meson-charmed meson scattering processes near the
threshold. In these reactions, $c$ and $\bar{c}$ are recombined into
a charmonium state, which is governed by the short range
interaction. To describe these meson-meson scatterings at the quark
level, we will employ the Barnes-Swanson quark-interchange model to
estimate the transition
amplitudes~\cite{Barnes:1991em,Barnes:1992qa,Wong:1999zb,Barnes:2000hu,Wong:2001td,Barnes:2003dg}.
In this approach, the non-relativistic quark potential model is
used, and the hadron-hadron scattering amplitudes are evaluated at
Born order with the interquark Hamiltonian. In the case of the
anticharmed meson-charmed meson scatterings, the amplitudes arise
from the sum of the four quark exchange diagrams as illustrated in
Fig. \ref{priordiagram}.

The interaction $H_{ij}$ between constituents $i$ and $j$ is
represented by the curly line in Fig. \ref{priordiagram}, and is
taken to be
\begin{eqnarray}
  H_{ij} &\equiv& \frac{\mathbf{\lambda}(i)}{2}\cdot\frac{\mathbf{\lambda}(j)}{2} V_{ij}({r_{ij}})  = \frac{\mathbf{\lambda}(i)}{2}\cdot\frac{\mathbf{\lambda}(j)}{2} \left( V_{{conf}}+V_{{hyp}} +V_{con} \right) \nonumber \\
  &=& \frac{\mathbf{\lambda}(i)}{2}\cdot\frac{\mathbf{\lambda}(j)}{2} \bigg\{
  \frac{\alpha_s}{r_{ij}}-\frac{3b}{4} r_{ij} -\frac{8\pi\alpha_s}{3m_i m_j}\mathbf{S}_i \cdot \mathbf{S}_j\  \left( \frac{\sigma^3}{\pi^{3/2}} \right) e^{-\sigma^2 r_{ij}^2} +V_{con} \bigg\}, \label{hamiltonian}
\end{eqnarray}
where for antiquarks the color Gell-Mann matrix $\mathbf{\lambda}$
should be replaced by $-{\mathbf{\lambda}}^T$. This Hamiltonian
contains a Coulomb-plus-linear confining potential $V_{conf}$ and a
short range spin-spin hyperfine term $V_{hyp}$, which is motivated
by one gluon exchange. The model parameters employed here are
$m_u$$=$$m_d$$=$$0.334$ GeV, $m_c$$=$$1.776$ GeV, $m_b$$=$$5.102$
GeV, $\sigma$$=$0.897 GeV, $b$$=$0.18 GeV$^2$, and the constant
$V_{con}$ is taken to be 0.62 GeV~\cite{Wong:2001td}. We use a
running coupling constant $\alpha_s(Q^2)$, which is given by
\begin{eqnarray}
\alpha_s(Q^2)=\frac{12\pi}{(33-2n_f)\mbox{ln}(A+Q^2/B^2)}
\end{eqnarray}
with $A$$=$10, $B$$=$0.31 GeV. And the scale $Q^2$ is identified
with the square of the invariant mass of the interacting
constituents. These conventional quark model parameters lead to a
good description of the meson spectrum, as listed in
Table~\ref{massspectrum}.

\begin{table}[htbp]
  \caption{Estimated meson masses according to the quark model. \label{massspectrum}}
\begin{center}
\begin{tabular}{|c|c|c|c|c|c|c|c|c|c|c|c|c|}
  \hline
   & $D$ & $D^*$ &$D(2S)$ & $D^*(2S)$ & $J/\psi$ & $\psi(2S)$ & $B$ & $B^*$ &$B_1$ & $\Upsilon(1S)$ & $\Upsilon(2S)$ & $\Upsilon(3S)$ \\
   \hline
  $M_{th}$ [GeV] & 1.920 & 1.993 & 2.711 & 2.769 & 3.089 & 3.701 & 5.387 & 5.411 & 5.748 & 9.471 &9.944 &10.347  \\
  \hline
  $M_{exp}$ [GeV] & 1.870 & 2.010 & 2.539 & 2.612 & 3.097 & 3.686 & 5.279 & 5.325 & 5.724 & 9.460  & 10.023 & 10.355  \\
  \hline
\end{tabular}
\end{center}
\end{table}

Following
Refs.~\cite{Barnes:1991em,Barnes:1992qa,Wong:1999zb,Barnes:2000hu,Wong:2001td,Barnes:2003dg},
we can write out the Born-order $T$-matrix element $T_{fi}$ which is the product of three
factors for each of the diagrams in Fig.\ref{priordiagram}, i.e.,
\begin{eqnarray}
T_{fi}=(2\pi)^3 I_{flavor}I_{color} I_{spin-space}.
\end{eqnarray}
Since there is no orbitally excited state involved in our
discussion, the factor $I_{spin-space}$ can be further factored into
\begin{equation}
I_{spin-space}=I_{spin} \times I_{space}.
\end{equation}
For the processes discussed in this paper, all of the quarks are
distinguishable, and we assume the external flavor states all have a
positive phase. The flavor factor $I_{flavor}$ is then simply unity.
For the color factor $I_{color}$, the diagrams C1 and C2 of Fig.
\ref{priordiagram} have factors of $-4/9$, and the diagrams T1 and
T2 have factors of $+4/9$. We list the spin factors for the operator
$\mathbf{S}_i \cdot \mathbf{S}_j$ of each diagram in
Table~\ref{SSspinfactor}. The spin factors of the unit operator for
all of the diagrams are equal, as listed in Table~\ref{Ispinfactor}.

\begin{table}[htbp]
  \caption{Spin factors $I_{spin}$ for the operator $\mathbf{S}_i \cdot \mathbf{S}_j$ of each diagram. Here, $S_A$ ($S_B$, $S_C$, $S_D$) is the spin of meson $A$ ($B$, $C$, $D$) in the reaction $AB\to CD$.\label{SSspinfactor}}
\begin{center}
\begin{tabular}{|c|c|c|c|c|c|c|c|c|}
  \hline
   ($S_A$,$S_B$)$\to$($S_C$,$S_D$)& C1 prior & C1 post & C2 prior & C2 post & T1 prior & T1 post & T2 prior & T2 post \\
   \hline
  ($0$, $1$)$\to$($0$, $1$) & $-\frac{3}{8}$ & $-\frac{3}{8}$ & $\frac{1}{8}$ & $\frac{1}{8}$ & $-\frac{1}{8}$ & $-\frac{1}{8}$ & $\frac{3}{8}$ & $\frac{3}{8}$ \\
  \hline
  ($1$, $1$)$\to$($0$, $1$) & $-\frac{3}{4\sqrt{2}}$ & $\frac{1}{4\sqrt{2}}$ & $\frac{1}{4\sqrt{2}}$ & $\frac{1}{4\sqrt{2}}$ & $-\frac{1}{4\sqrt{2}}$ & $-\frac{1}{4\sqrt{2}}$ &$-\frac{1}{4\sqrt{2}}$ &$-\frac{1}{4\sqrt{2}}$ \\
  \hline
\end{tabular}
\end{center}
\end{table}

\begin{table}[htbp]
  \caption{Spin factors $I_{spin}$ for the unit operator.\label{Ispinfactor}}
\begin{center}
\begin{tabular}{|c|c|}
  \hline
  ($S_A$,$S_B$)$\to$($S_C$,$S_D$) & All diagrams \\
  \hline
  ($0$, $1$)$\to$($0$, $1$) & $\frac{1}{2}$ \\
  \hline
  ($1$, $1$)$\to$($0$, $1$) & $\frac{1}{\sqrt{2}}$ \\
  \hline
\end{tabular}
\end{center}
\end{table}

The space factors are evaluated as the overlap integrals of the
quark model wave functions. It is convenient to write these overlap
integrals in the momentum-space. For the four diagrams of
Fig.~\ref{priordiagram}, in the reaction $AB\to CD$, where $AB$ and
$CD$ are the initial and final meson pairs, respectively. The space
factors read as
\begin{eqnarray}
&& I_{space}^{C1-prior} = \int\int  d\mathbf{k}\ d\mathbf{q}\  \Phi_A(2\mathbf{k})\ \Phi_B(2\mathbf{k}-2\mathbf{P_C})\ \Phi_C(2\mathbf{q}-\mathbf{P_C})\ \Phi_D(2\mathbf{k}-\mathbf{P_C}) \ V(\mathbf{k}-\mathbf{q}),  \\
&& I_{space}^{C2-prior} = \int\int  d\mathbf{k}\ d\mathbf{q}\  \Phi_A(-2\mathbf{k})\ \Phi_B(-2\mathbf{k}-2\mathbf{P_C})\ \Phi_C(-2\mathbf{k}-\mathbf{P_C})\ \Phi_D(-2\mathbf{q}-\mathbf{P_C}) \ V(\mathbf{k}-\mathbf{q}),   \\
&& I_{space}^{T1} = \int\int  d\mathbf{k}\ d\mathbf{q}\  \Phi_A(2\mathbf{k})\ \Phi_B(2\mathbf{q}-2\mathbf{P_C})\ \Phi_C(2\mathbf{q}-\mathbf{P_C})\ \Phi_D(2\mathbf{k}-\mathbf{P_C}) \ V(\mathbf{k}-\mathbf{q}),  \\
&& I_{space}^{T2} = \int\int  d\mathbf{k}\ d\mathbf{q}\
\Phi_A(-2\mathbf{k})\ \Phi_B(-2\mathbf{q}-2\mathbf{P_C})\
\Phi_C(-2\mathbf{k}-\mathbf{P_C})\ \Phi_D(-2\mathbf{q}-\mathbf{P_C})
\ V(\mathbf{k}-\mathbf{q}), \label{eqprior}\end{eqnarray} where
$\bf{P_C}$ is the center-of mass momentum of meson $C$. And the
potential $V(\mathbf{p})$ is obtained via the Fourier transformation of
$V(r)$. Taking into account that, we are actually estimating the
decaying amplitudes of the molecular states, of which the components
are $A$ and $B$. Therefore, the center-of-mass momenta of $A$ and $B$
can be approximately taken as
$\mathbf{P_A}$$=$$-\mathbf{P_B}$$\approx$$0$. We have used these
approximations in deriving the above equations.

\begin{figure}[tb]
  \centering
  \includegraphics[width=0.4\hsize]{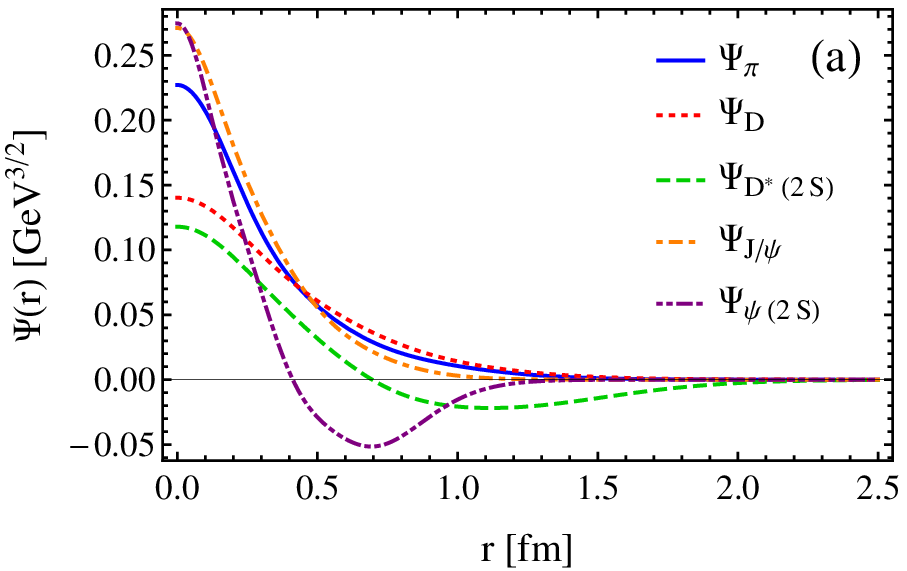}\\ \includegraphics[width=0.4\hsize]{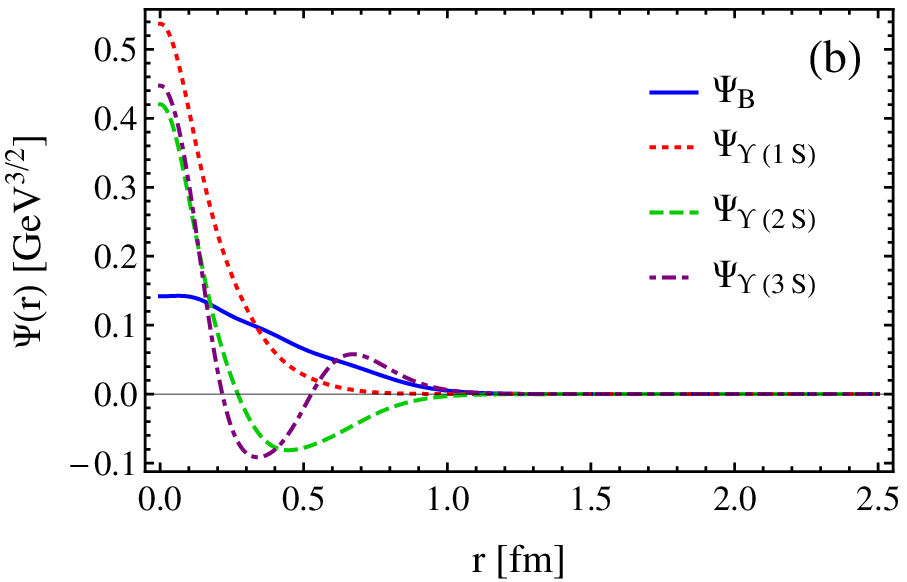} \\ \includegraphics[width=0.4\hsize]{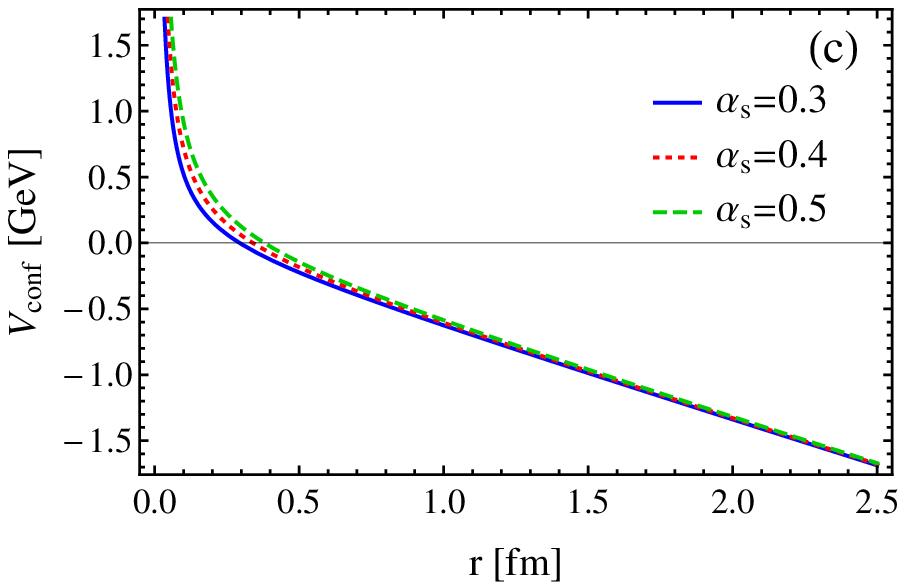}\\
  \caption{(a) Wave functions of $\pi$, $D$, $D^{*}(2S)$, $J/\psi$ and $\psi (2S)$ in coordinate space. (b) Wave functions of $B$, $\Upsilon(1S)$, $\Upsilon(2S)$ and $\Upsilon(3S)$ in coordinate space. (c) the r-dependence of the confining potential $V_{conf}$$=$$\frac{\alpha_s}{r} - \frac{3b}{4}r$ with $b$=0.18 GeV$^2$.}\label{wavefunction}
\end{figure}

To simplify the spatial overlap integrals, we expand the wave
functions $\Psi(\mathbf{r})$ in the coordinate space of each state
as a linear combination of Gaussian basis
functions~\cite{Wong:2001td}, i.e.
\begin{eqnarray}
\Psi(\mathbf{r})=\sum_{n=1}^N a_n \
\frac{(n\beta^2)^{3/4}}{\pi^{3/4}}\ e^{-n\beta^2 \mathbf{r}^2/{2}},
\end{eqnarray}
where $\beta$ denotes the width parameter of Gaussian functions. $a_n$ is
the expansion coefficient, and we take $N$=6 in our calculations.
Using the interaction in Eq.~(\ref{hamiltonian}) and solving the
eigenvalue equation, we obtain the mass spectrum and the corresponding wave functions
which are displayed in Table~\ref{massspectrum} and
Fig.~\ref{wavefunction}, respectively. {To fit the pertinent hadron mass spectrum well, we set the width parameter $\beta$ as 0.3 GeV and 0.61 GeV for charmed mesons and bottomed mesons, respectively, which is different from the fitting parameters in Ref~\cite{Wong:2001td}.} Since the wave functions of
the radially excited states have nodes, the overlap integrals for
$\pi J/\psi$ and $\pi\psi(2S)$ final states are very different. In
other words, the branching fractions of $Z_c(4430)$ ($Z_c(3900)$, and
$Z_c(4020)$) decaying into $\pi J/\psi$ and $\pi \psi(2S)$ will be
different.

There is still a "prior-post" ambiguity while calculating the
scattering amplitudes via the quark-interchange model
~\cite{Barnes:1991em,Barnes:1992qa,Wong:1999zb,Barnes:2000hu,Wong:2001td,Barnes:2003dg}.
The Hamiltonian that describes the $AB\to CD$ process can be
separated into free and interaction parts in two ways, i.e.
$H$$=$$H^{0}_A$$+$$H^{0}_B$$+$$V_{AB}$ or
$H$$=$$H^{0}_C$$+$$H^{0}_D$$+$$V_{CD}$. In the "prior" scattering
diagrams of Fig.~\ref{priordiagram}, the interactions occur before
the quark interchange, and the interaction Hamiltonian is taken as
$V_{AB}$. There are also some corresponding "post" diagrams
displayed in Fig.~\ref{postdiagram}, where the interactions occur
after the quark interchange, and the interaction Hamiltonian is
$V_{CD}$. The spin factors for the post diagrams are also listed in
Tables~\ref{SSspinfactor} and \ref{Ispinfactor}. The space factors
for $T_1$ and $T_2$ are the same as those of prior diagrams, and
the factors for C1-post and C2-post read as
\begin{eqnarray}
I_{space}^{C1-post} &=& \int\int  d\mathbf{k}\ d\mathbf{q}\  \Phi_A(2\mathbf{k})\ \Phi_B(2\mathbf{q}-2\mathbf{P_C})\ \Phi_C(2\mathbf{q}-\mathbf{P_C})\ \Phi_D(2\mathbf{q}-\mathbf{P_C}) \ V(\mathbf{k}-\mathbf{q}), \\
I_{space}^{C2-post} &=& \int\int  d\mathbf{k}\ d\mathbf{q}\
\Phi_A(-2\mathbf{k})\ \Phi_B(-2\mathbf{q}-2\mathbf{P_C})\
\Phi_C(-2\mathbf{k}-\mathbf{P_C})\ \Phi_D(-2\mathbf{k}-\mathbf{P_C})
\ V(\mathbf{k}-\mathbf{q}).
\end{eqnarray}
If one uses the exact bound state wave functions determined by the
full Hamiltonian, and the interaction used in calculating the
scattering amplitudes is identical to that used in calculating the
wave functions, the prior and post scattering amplitudes should be
equal. However, the numerical test of the prior-post equivalence
shows some relatively large sensitivities to the wave functions
~\cite{Barnes:1991em,Barnes:1992qa,Wong:1999zb,Barnes:2000hu,Wong:2001td,Barnes:2003dg}.

In the present case, this ambiguity can be greatly removed based on
the physical considerations. According to the OZI rule, the
interaction between the charmed meson and anti-charmed meson should
be much stronger than that between the light hadron and charmonium
state, which was confirmed by the lattice QCD
simulation~\cite{Liu:2008rza}. Taking into account this point, we
will mainly discuss the scattering process according to the "prior"
diagrams.

\begin{figure}[tb]
  \centering
  \includegraphics[width=0.33\hsize]{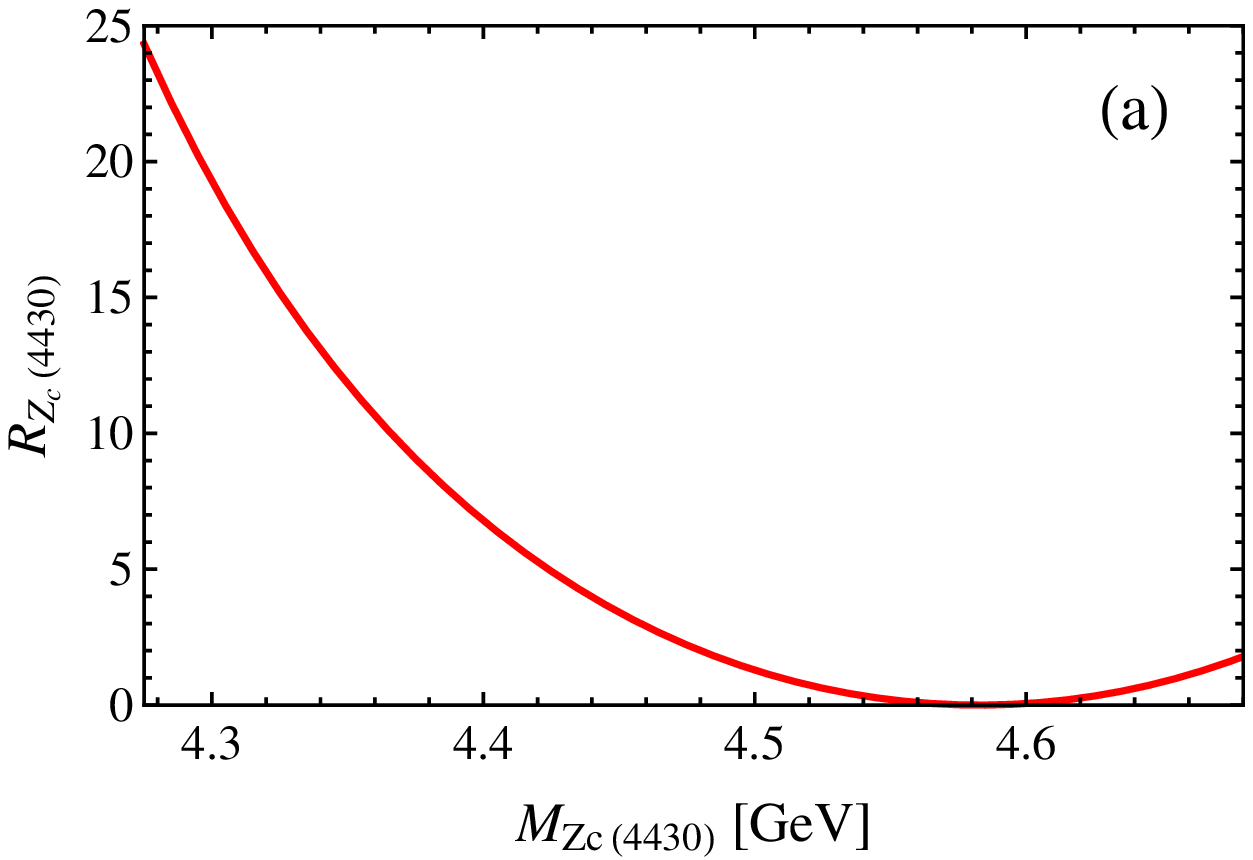}\includegraphics[width=0.326\hsize]{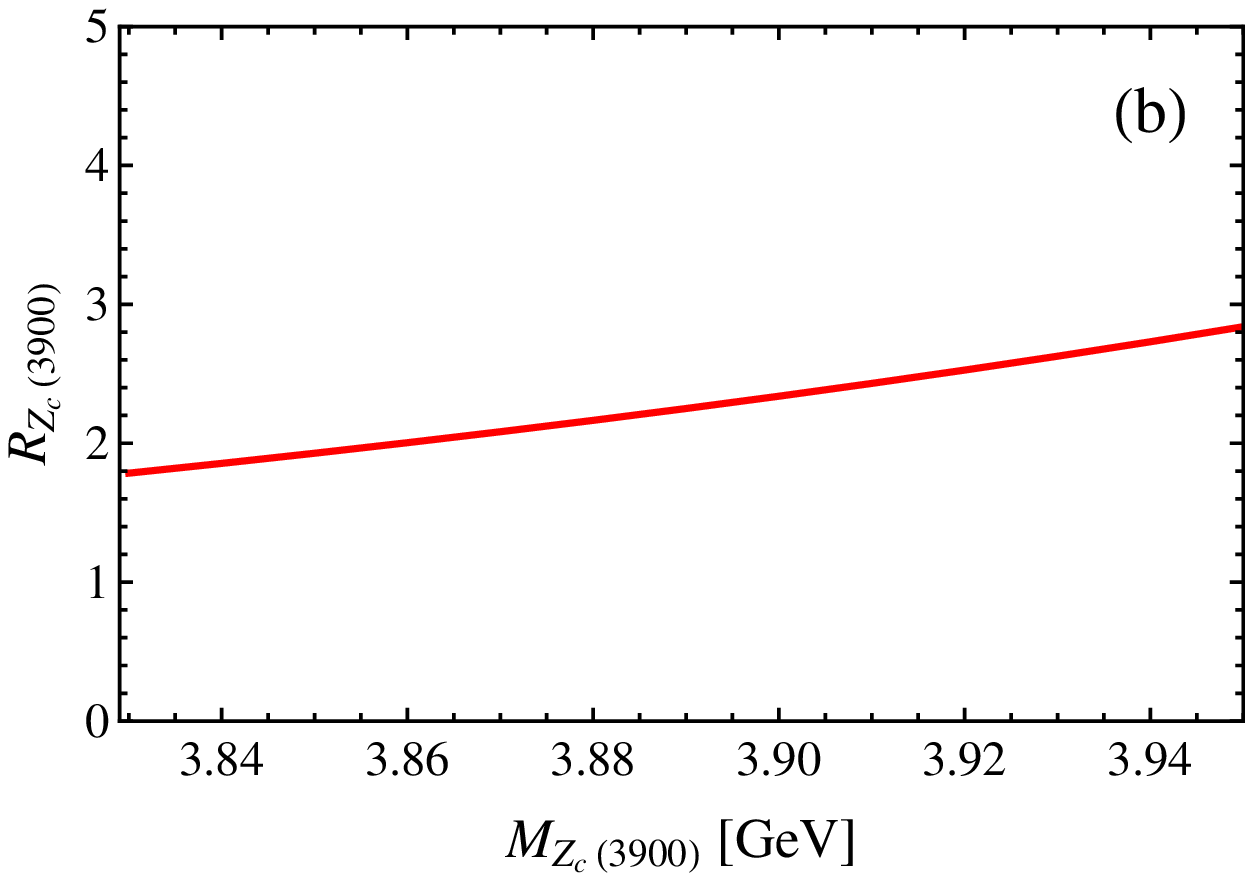}\includegraphics[width=0.34\hsize]{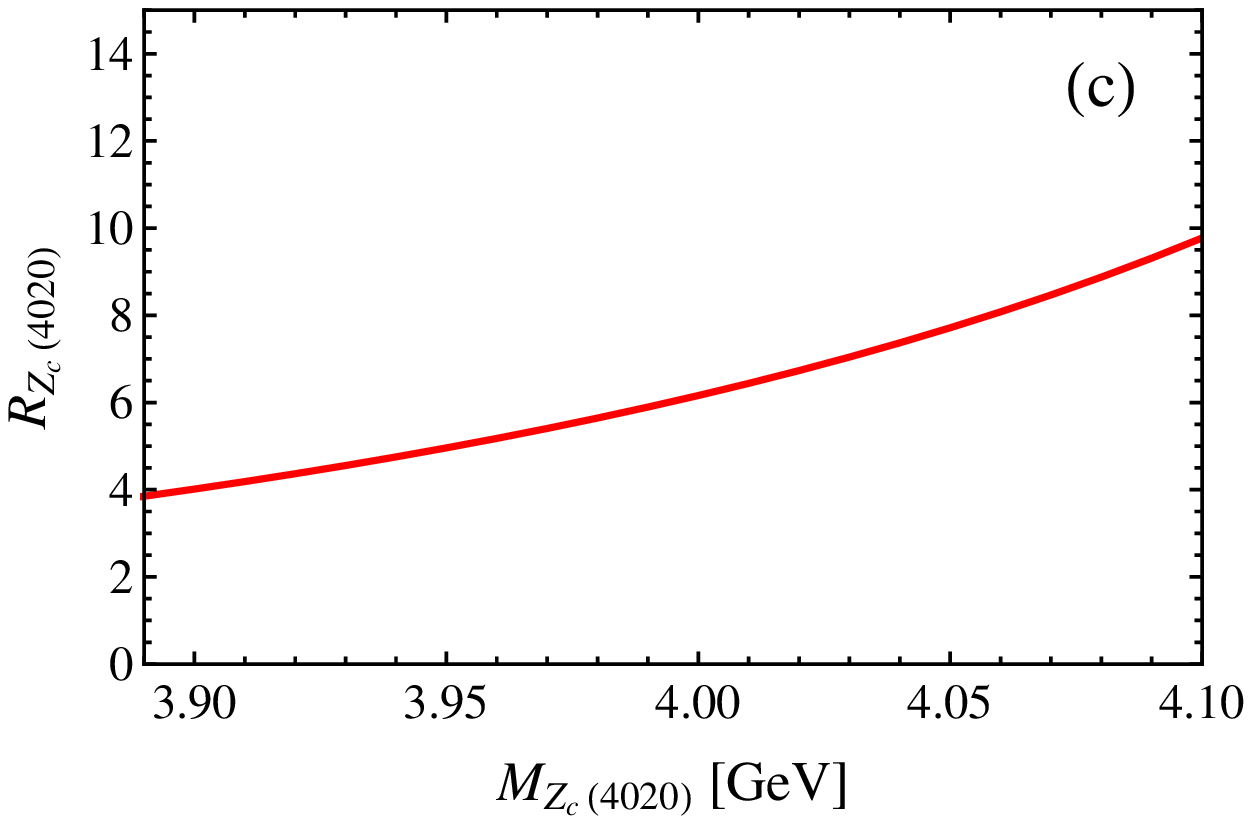}\\
  \caption{The mass-dependence of the ratios $R_{Z_c(4430)}$, $R_{Z_c(3900)}$ and $R_{Z_c(4020)}$.}\label{Zcratio}
\end{figure}

\begin{figure}[tb]
  \centering
  \includegraphics[width=0.35\hsize]{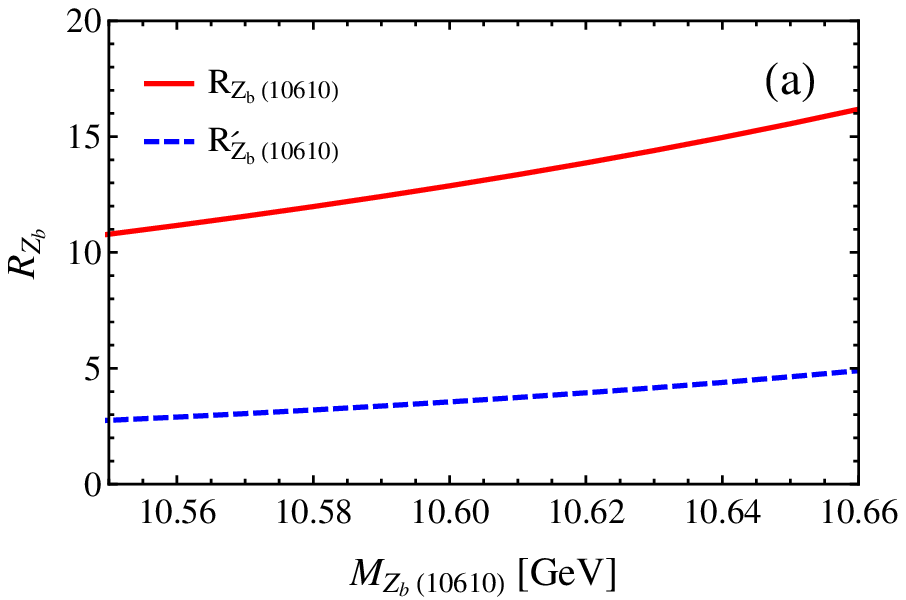}\includegraphics[width=0.35\hsize]{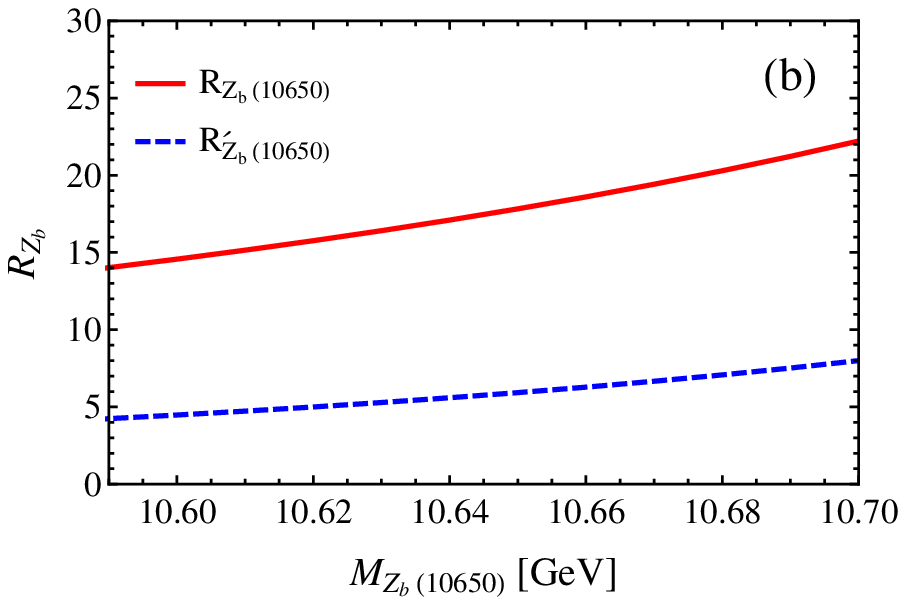}\\
  \caption{The mass-dependence of the ratios $R_{Z_b(10610)}$, $R_{Z_b(10610)}^{\prime}$, $R_{Z_b(10650)}$ and $R_{Z_b(10650)}^{\prime}$.}\label{Zbratio}
\end{figure}

\section{Numerical Results}

In this work, we focus on the difference between the branching
fractions of $Z_c^\pm$ ($Z_b^\pm$) decaying into $J/\psi\pi^\pm$
($\Upsilon(1S)\pi^\pm$) and $\psi(2S)\pi^\pm$
($\Upsilon(2S,3S)\pi^\pm$), and ignore the dynamics of the heavy
meson pairs forming the molecular states for the moment. We
introduce the ratios of the branching fractions of the XYZ state to
the radially excited and ground states as follows:
 \begin{eqnarray}
 R_{Z_c}\equiv \frac{\Gamma [Z_c\to \pi\psi(2S) ]} {\Gamma [ Z_c\to \pi J/\psi]},\ \ R_{Z_b}\equiv \frac{\Gamma [Z_b\to \pi\Upsilon(2S) ]} {\Gamma [ Z_b\to \pi \Upsilon(1S)]},\ \ R_{Z_b}^\prime \equiv \frac{\Gamma [Z_b\to \pi\Upsilon(3S) ]} {\Gamma [ Z_b\to \pi \Upsilon(1S)]},
 \label{ratios}\end{eqnarray}
where $Z_c$ could be $Z_c(4430)$, $Z_c(3900)$ and $Z_c(4020)$
respectively, and $Z_b$ could be $Z_b(10610)$ and $Z_b(10650)$
respectively. Since the measured masses of these molecular
candidates still have relatively large uncertainties, we will first
give the mass dependence of the ratios in Eq.~(\ref{ratios}). The
numerical results are displayed in Figs.~\ref{Zcratio} and
\ref{Zbratio}. Apart from $R_{Z_c(4430)}$, the other ratios are not
sensitive to the masses of the corresponding molecular states.


Some of these molecular candidates have large width. For instance,
the width of $Z_c(4430)$ is about 172 MeV~\cite{Aaij:2014jqa}. One
therefore has to take into account the mass distribution of these
states while investigating their decay channels. Their two-body
decay width can be calculated as:
\begin{eqnarray}
\Gamma[Z_c]_{\mbox{2-body}} =\frac{1}{W}
\int^{(m_{Z_c}+2\Gamma_{Z_c})^2}_{s_{th}} ds\
\frac{(2\pi)^4}{2\sqrt{s}} \int d\Phi_2\ |\mathcal{M}_{fi}|^2
\frac{1}{\pi} \mbox{Im}\left( \frac{-1}{s-m_{Z_c}^2+i m_{Z_c}
\Gamma_{Z_c}}  \right),
\end{eqnarray}
where $d\Phi_2$ is the two-body phase space. $s_{th}$ is the energy
threshold, and we set $s_{th}$ as $(m_{\psi(2S)}$+$m_\pi)^2$ and
$(m_{\Upsilon(3S)}$+$m_\pi)^2$ for $Z_c$ and $Z_b$ decays
respectively. $W$ is a normalization factor, which reads as
\begin{eqnarray}
W=\frac{1}{\pi} \int^{(m_{Z_c}+2\Gamma_{Z_c})^2}_{s_{th}} ds\
\mbox{Im}\left( \frac{-1}{s-m_{Z_c}^2+i m_{Z_c} \Gamma_{Z_c}}
\right).
\end{eqnarray}
$\mathcal{M}_{fi}$ is related to $T_{fi}$ by
\begin{eqnarray}
\mathcal{M}_{fi}=\sqrt{E_A E_B E_C E_D}\,  T_{fi},
\end{eqnarray}
where $E_A$, $E_B$, $E_C$ and $E_D$ are the energies of meson $A$,
$B$, $C$ and $D$ in the center-of-mass system, respectively, for
$AB$$\to$$CD$ reaction. The obtained ratios are listed in
Table~\ref{widthratio}, where we have used the center values of
experimental data for $M_{Z_c}$, $M_{Z_b}$, $\Gamma_{Z_c}$ and
$\Gamma_{Z_b}$.

Assuming that the components of $Z_c(4430)$ are composed of
$\bar{D}D^*(2S)$ \cite{Ma:2014zua}, we obtain
$R_{Z_c(4430)}$$\approx$9.8, which is very close to the experimental
data. The nodes in the wave functions of $D^*(2S)$ and $\psi(2S)$
ensure the larger spatial overlap integral and render $\pi \psi(2S)$
to be a more favorable mode of $Z_c(4430)$.

If $Z_c(3900)$ and $Z_c(4020)$ are the $\bar{D}D^*$ and
$\bar{D}^*D^*$ molecular states, respectively, \cite{LX-1}, we derive the ratios
$R_{Z_c(3900)}$$\approx$1.3 and $R_{Z_c(4020)}$$\approx$$4.7$.
According to the above results, it seems that these molecular states
are more likely to decay into $\pi\psi(2S)$ than into $\pi J/\psi$.
This conclusion can be qualitatively understood according to the
node positions of the wave functions and potentials. In
Fig.~\ref{wavefunction}, we also display the r-dependence of the
confining potential $V_{con}$, which crosses zero around 0.4 fm. The
wave function of $\psi(2S)$ also has a node around 0.4 fm while the
wave function of $J/\psi$ does not have a node. Recall that the wave
functions of the initial states are the same and do not have nodes.
Therefore the overlap integral for $\pi J/\psi$ final states is much
smaller than that for $\pi\psi(2S)$ final states. Notice that
$R_{Z_c(4020)}$ is much larger than 1, which implies that it is very
hopeful to find $Z_c(4020)$ in the final states containing
$\pi\psi(2S)$, such as the reaction $e^+e^-$$\to$$\psi(2S)\pi\pi$.

Similarly, we assume $Z_b(10610)$ and $Z_b(10650)$ to be molecular
states composed of $B\bar{B}^*$ and $B^*\bar{B}^*$ respectively
\cite{LX-2}. As displayed in Table~\ref{widthratio}, the estimated
ratios for $Z_b(10610)$ and $Z_b(10650)$ are also very close to the
experimental measurements by the Belle Collaboration, which favors the
$B\bar{B}^*$ and $B^*\bar{B}^*$ molecular states explanations
concerning the $Z_b$ states.

Although some results which are consistent with the data are
obtained, we should mention that there are still some theoretical
uncertainties within the present framework. One major uncertainty
arises from the hadron wave functions, because it may not be a very
good approximation to treat the pion meson and open flavor mesons as the
non-relativistic systems. However, the potentials and wave functions
do not vary too much among different quark models. Our qualitative
estimation of the overlap integrals mentioned in the last several paragraphs
still works, which will be a less model-dependent conclusion.

\begin{table}[htbp]
\caption{The ratios after taking into account the width of the molecular
candidate. Here, the experimental ratios are estimated according to
Refs.~\cite{shencp:2014,Garmash:2014dhx}, and we only take into
account the statistical error of the data. \label{widthratio}}
\begin{center}
\begin{tabular}{|c|c|c|c|c|c|c|c|}
  \hline
   & $R_{Z_c(4430)}$ & $R_{Z_c(3900)}$ & $R_{Z_c(4020)}$ & $R_{Z_b(10610)}$ & $R^\prime_{Z_b(10610)}$ & $R_{Z_b(10650)}$ & $R^\prime_{Z_b(10650)}$ \\
   \hline
  Theo. & 9.8 & 1.3 & 4.7 & 11.4 & 2.2 & 15.8 & 4.0 \\
  \hline
  Exp. & $\sim$10 & $\cdots$ & $\cdots$ &$4.8\sim 8.8$ & $2.7\sim 5.3$  & $4.5\sim 12.1$ &$5.4\sim 14.0$  \\
  \hline
\end{tabular}
 \end{center}
\end{table}

\section{Summary}

In this work, we have systematically discussed the puzzling decay
properties of the exotic states $Z_c(3900)$, $Z_c(4020)$, and their
radially excited cousin $Z_c(4430)$. With the quark-interchange
model, we have derived the ratios of the branching fractions of
these molecular candidates decaying into the ground states and
radially excited states. Our numerical results imply that these
molecular candidates are more likely to decay into radially excited
states than into ground states.

As a $\bar{D}D^*(2S)$ molecular candidate, $R_{Z_c(4430)}$ tends to
decay into $\pi\psi(2S)$ more easily. The interplay of the node in the
wave function of $D^*(2S)$ and the node in the wave function of
$\psi(2S)$ ensures the larger spatial overlap integral of the
scattering amplitude. We obtain the ratio $R_{Z_c(4430)}\sim 9.8$,
which is close to the experimental measurement. In fact, the
measured ratio favors the $\bar{D}D^*(2S)$ molecular state ansatz
concerning the intrinsic structure of $Z_c(4430)$. The above decay
pattern is very characteristic for a molecular state containing a
radially excited component.

In contrast, if $Z_c(4430)$ is an S-wave tetraquark ground state, it
will be extremely challenging to accommodate the experimental ratio
$R_{Z_c(4430)}\sim 10$. If $Z_c(4430)$ is a radially excited
tetraquark state, one may also expect $\pi\psi(2S)$ to be a more
favorable decay mode. Then, where is the ground state tetraquark?

We have also estimated the ratios for $Z_b(10610)$ and $Z_b(10650)$
assuming that they are molecular candidates. Our numerical results show
that the favorable decay modes of $Z_b(10610)$ and $Z_b(10650)$ are
$\pi\Upsilon(2S)$ and $\pi\Upsilon(3S)$, which is consistent with
the experimental measurement by the Belle Collaboration.

Assuming $Z_c(3900)$ and $Z_c(4020)$ are the $\bar{D}D^*$ and
$\bar{D}^*D^*$ molecular states, we have predicted the ratios
$R_{Z_c(3900)}$$\approx$1.3 and $R_{Z_c(4020)}$$\approx$$4.7$. The
accidental match of the node of the interaction potential with the
node of the wave function of $\psi(2S)$ enhances the spatial overlap
integral of the scattering amplitude and renders $\pi\psi(2S)$ to be a
favorable decay mode. Hopefully the $\pi\psi(2S)$ mode, and ratios
$R_{Z_c(3900)}$ and $R_{Z_c(4020)}$ will be measured by the BESIII
and Belle collaborations in the near future, which shall be very
helpful to understand the underlying dynamics of these exotic
states.

\subsection*{Acknowledgments}
We would like to thank the helpful discussion with C.~P. Shen. This work was supported by the China Postdoctoral Science Foundation
under Grant No. 2013M530461, the National Natural Science Foundation
of China under Grants No. 11222547, No. 11175073, No. 11035006, No.
11375240 and No. 11261130311, the Ministry of Education of China
(FANEDD under Grant No. 200924, SRFDP under Grant No. 2012021111000,
and NCET), the Fok Ying Tung Education Foundation (Grant No.
131006), and the DFG and the NSFC through funds provided to the
Sino-German CRC 110 "Symmetries and the Emergence of Structure in
QCD".

\end{document}